\documentstyle[12pt,aaspp4,epsfig]{article}
\begin{document}

\title{Spin Down of Pulsations in the Cooling Tail of an X-ray Burst from 4U 1636-53}
\author{Tod E. Strohmayer}
\affil{Laboratory for High Energy Astrophysics, NASA's Goddard Space Flight 
Center, Greenbelt, MD 20771; stroh@clarence.gsfc.nasa.gov}
\authoraddr{Laboratory for High Energy Astrophysics, Mail Code 662, NASA/GSFC
Greenbelt, MD 20771}

\begin{abstract}

We report the discovery with the proportional counter array (PCA) onboard the 
Rossi X-ray Timing Explorer (RXTE) of a decrease in the frequency of X-ray brightness 
oscillations in the cooling tail of an X-ray burst from 4U 1636-53. This
is the first direct evidence for a spin down of the pulsations seen during 
thermonuclear bursts.  We find that the spin down episode 
is correlated with the appearance in this burst of an extended tail of emission with a 
decay timescale much longer than is seen in other bursts from 4U 1636-53 in the same 
set of observations. We present both time resolved energy and variability spectra
during this burst and compare them with results from a second burst which shows
neither a spin down episode nor an extended tail.
A spectral evolution study of the ``spin down'' burst reveals a secondary signature of 
weak radius expansion, not seen in other bursts, and correlated with the spin down 
episode, which may indicate a secondary thermonuclear energy release. 
We interpret the spin down episode in the context of an angular momentum conserving 
shell, which is reexpanded and therefore spun down by an additional thermonuclear 
energy release which could also explain the extended X-ray tail. 

\end{abstract}


\centerline{Accepted for Publication in the Astrophysical Journal Letters}

\vfill\eject

\section{Introduction}

Millisecond oscillations in the X-ray brightness during thermonuclear bursts,
``burst oscillations", have been observed from six low
mass X-ray binaries (LMXB) with the Rossi X-ray Timing Explorer (RXTE) (see
\markcite{SSZ}Strohmayer, Swank \& Zhang {\it et al.} 1998 for a recent review).
Considerable evidence points to rotational modulation as the source of these
pulsations (see for example, \markcite{SZS}Strohmayer, Zhang \& Swank 1997; 
\markcite{SM99}Strohmayer \& Markwardt 1999). Anisotropic X-ray emission caused by 
either localized or
inhomogeneous nuclear burning produces either one or a pair of hot spots on the
surface which are then modulated by rotation of the neutron star. A remarkable 
property of these oscillations is the frequency evolution which occurs in the 
cooling tail of some bursts. Recently, \markcite{SM99}Strohmayer \& Markwardt (1999) 
have shown that the frequency in the cooling tail of bursts from 4U 1728-34 and 
4U 1702-429 is well described by an exponential chirp model whose frequency increases 
asymptotically toward a limiting value. \markcite{SJGL}Strohmayer et. al (1997) 
have argued this evolution results from angular momentum conservation of the 
thermonuclear shell, which cools, shrinks and spins up as the surface radiates 
away the thermonuclear energy. To date, only frequency increases have been reported
in the cooling tails of bursts, consistent with settling of the shell as its energy is
radiated away.

In this Letter we report observations of a {\it decreasing} burst oscillation
frequency in the tail of an X-ray burst. We find that an episode of spin down in
the cooling tail of a burst observed on December 31, 1996 at 17:36:52 UTC 
(hereafter, burst A, or the ``spin down'' burst)
from 4U 1636-53 is correlated with the presence of an extended. In \S 1 we present 
an analysis of the frequency evolution in this burst, with emphasis on the spin down 
episode. In \S 2 we present time resolved energy spectra of the spin down burst, and we 
investigate the energetics of the extended tail. Throughout, we compare the temporal 
and spectral behavior of the spin down burst with a different burst observed on 
December 29, 1996 at 23:26:46 UTC (hereafter, burst B) which does not show either a 
spin down episode nor an extended tail of emission, but which is similar to the spin 
down burst in most other respects. We conclude in \S 3 with a summary and discussion of 
the spin down episode and extended emission in the context of an additional, delayed 
thermonuclear energy release which might re-expand the thermonuclear shell and perhaps 
account for both the spin down and the extended tail of thermal emission.

\section{Evidence for Spin Down}

Oscillations at 580 Hz were discovered in thermonuclear bursts from
4U 1636-53 by Zhang et al. (1996). More recently, Miller (1999a) has reported evidence
during the rising phase of bursts of a significant modulation at half the 580 Hz 
frequency  suggesting that 580 Hz is twice the neutron star spin frequency
and that a pair of antipodal spots produce the oscillations. Here we focus on a 
burst from 4U 1636-53 which shows a unique decrease in the $\approx 580$ Hz oscillation 
frequency. To study the evolution in frequency of burst oscillations we employ the 
$Z_n^2$ statistic (\markcite{B83}Buccheri et al. 1983). We have described this 
method previously, and details can be found in \markcite{SM99}Strohmayer \& 
Markwardt (1999). We first 
constructed for both bursts A and B a dynamic ``variability'' spectrum by computing 
$Z_1^2$ as a function of time on a grid of frequency values in the vicinity of 580 Hz. 
We used 2 second intervals to compute $Z_1^2$ and started a new interval every 0.25 
seconds. This variability spectrum is very similar to a standard dynamic power spectrum,
however, the $Z_1^2$ statistic allows for a more densely sampled frequency grid than a 
standard Fast Fourier Transform power spectrum. The results are shown in Figure 1 
(bursts A and B are in the top and bottom panels, respectively) as contour maps of 
constant $Z_1^2$ through each burst. The contour map for the spin down burst (top 
panel) suggests that the oscillation began with a frequency near 579.6 Hz at burst 
onset, reappeared later in the burst after ``touchdown'' of the photosphere at an 
increased frequency, $\approx 580.7$ Hz, but then beginning near 11 seconds dropped to 
$\approx 579.6$ Hz over several seconds. For comparison, we also show in Figure 1 
(bottom panel) a similar variability spectrum for burst B which also shows strong 
oscillations near 580 Hz, but shows no evidence of a similar spin down episode.

To investigate the evolution of the oscillation frequency more quantitatively we fit 
a model for the temporal evolution of the frequency, $\nu(t)$, to the 4.5 second 
interval during which the oscillation is evident in the dynamic variability spectrum 
(Figure 1, top panel). Our model is composed of two linear segments, each with its 
own slope, joined continuously at a break time $t_b$. This is similar to the model 
employed by \markcite{M99b}Miller (1999b), and has four free parameters, the initial 
frequency, $\nu_0$, the two slopes, $d_{\nu}^1$ and $d_{\nu}^2$, and the break time, 
$t_b$. We used this frequency model to compute phases $\phi_{t_j}$ for each X-ray 
event, {\it viz.} $\phi_{t_j} = \int_0^{t_j} \nu(t') dt'$, where $t_j$ are the 
photon arrival times, and then varied the model parameters to maximize the $Z_1^2$ 
statistic. We used a downhill simplex method for the maximization (see Press et al. 
1989). Figure 2 compares $Z_1^2$ vs. parameter $\nu_0$ for the best fitting two 
segment model (solid histogram) and a simple constant frequency model ($\nu(t) = 
\nu_0$, dashed histogram). The two segment model produces a significant increase in the 
maximum $Z_1^2$ of about 40 compared with no frequency evolution, and it also 
yields a single, narrower peak in the $Z_1^2$ distribution.  The 
increase of 40 in $Z_1^2$, which for a purely random process is distributed as $\chi^2$ 
with 2 degrees of freedom, argues convincingly that the frequency drop is significant.
We note that \markcite{M99b}Miller (1999b) has also identified the same spin down 
episode during this burst using a different, but related method. The best fitting two 
segment model is shown graphically as the solid curve in Figure 1 (top panel). 

\section{Time history, spectral evolution and burst energetics}

A comparison of the 2 - 20 keV time history of the spin down burst with other bursts 
from the same observations reveals that this burst is also unique in having an
extended tail of thermal emission. This is well illustrated in Figure 3, 
which compares the 2 - 20 keV time histories of the spin down burst and burst B. 
To further investigate the energetics of the 
thermal burst emission we performed a spectral evolution analysis. 
We accumulated energy spectra for varying time intervals through both bursts. Using 
XSPEC we fit blackbody spectra to each interval by first subtracting a pre-burst interval 
as background, and then investigated the temporal evolution of the blackbody temperature, 
$kT$, inferred radius, $R_{BB}$ and bolometric flux, $F$. In most intervals we obtained 
acceptable fits with the blackbody model. The results for both bursts are summarized 
in Figure 4, We have aligned the burst profiles in time
for direct comparison. Both bursts show evidence for radius expansion shortly after 
onset in that $kT$ drops initially and then recovers. Their peak fluxes are also 
similar, consistent with being Eddington limited. Out to about 7 
seconds post-onset both bursts show the same qualitative behavior, after this, 
however, the spin down burst (solid curve) shows a much more gradual decrease 
in both the blackbody temperature $kT$ and the bolometric flux $F$ than is evident in 
burst B. We integrated the flux versus time profile for each burst in order to estimate 
fluences and establish the energy budget in the extended tail. We find fluences of $1.4 
\times 10^{-6}$ and $5.1 \times 10^{-7}$ ergs cm$^{-2}$ for bursts A and B 
respectively. That is, the spin down burst has about 2.75 times more energy than burst 
B. Put another way, most of the energy in the spin down burst is in the extended tail. 
In figure 4 we also indicate with a vertical dotted line the time $t_b$ associated with 
the beginning of the spin down episode based on our modelling of the 580 Hz 
oscillations. The spectral evolution analysis indicates that at about the same
time the spin down episode began there was also a change in its spectral evolution 
as compared with that of burst B. This behavior is evident in figure 5 which 
shows the evolution of $kT$ (dashed curve) and the inferred blackbody radius $R_{BB}$ 
(solid curve) for the spin down burst. Notice the secondary increase in $R_{BB}$ and an 
associated dip in $kT$ near time $t_b$ (vertical dotted line). This behavior 
is similar to the signature of radius expansion seen earlier in both bursts, but at
a weaker level, it suggests that at this time there may have been an additional 
thermonuclear energy input in the accreted layers, perhaps at greater depth, which 
then diffused out on a longer timescale, producing the extended tail. That this 
spectral signature ocurred near the same time as the onset of the spin down 
episode suggests that the two events may be causally related. 

\section{Discussion and Summary}

The observation of thermonuclear bursts with extended tails is not a new phenomenon. 
\markcite{CCG}Czerny, Czerny, \& Grindlay (1987) reported on a burst from the soft X-ray
transient Aql X-1 which showed a long, relatively flat X-ray tail. Bursts following 
this one were found to have much shorter, weaker tails. \markcite{F92}Fushiki et al. 
(1992) argued that such long tails were caused by an extended phase of hydrogen burning 
due to electron captures at high density ($\rho \approx 10^7$ g 
cm$^{-3}$) in the accreted envelope. Such behavior is made possible because of the long 
time required to accumulate an unstable pile of hydrogen-rich thermonuclear fuel when 
the neutron star is relatively cool ($\approx 10^7$ K) prior to the onset of accretion. 
This, they argued, could occur in transients such as Aql X-1, which have long quiescent 
periods during which the neutron star envelope can cool, and thus the first 
thermonuclear burst after the onset of an accretion driven outburst should show the 
longest extended tail. Other researchers have shown that the thermal state of the 
neutron star envelope, the abundance of CNO materials in the accreted matter, and 
variations in the mass accretion rate all can have profound effects on the character of 
bursts produced during an accretion driven outburst (see \markcite{T93}Taam et al. 
1993; \markcite{WW85}Woosley \& Weaver 1985; and \markcite{AJ82}Ayasli \& Joss 1982, 
and references therein). For example, \markcite{T93}Taam et al. (1993) showed that for 
low CNO abundances and cool neutron star envelopes the subsequent
bursting behavior can be extremely erratic, with burst recurrence times varying by as 
much as two orders of magnitude. They also showed that such conditions produce dwarf 
bursts, with short recurrence times and peak fluxes less than a tenth Eddington, and 
that many bursts do not burn all the fuel accumulated since the last X-ray burst. Thus 
residual fuel, in particular hydrogen, can survive and provide energy for subsequent
bursts. These effects lead to a great diversity in the properties of X-ray bursts
observed from a single accreting neutron star. Some of these effects were likely 
at work during the December, 1996 observations of 4U 1636-53 discussed here, as 
both a burst with a long extended tail, as well as a dwarf burst were observed (see 
\markcite{M99b}Miller 1999b).

The spin up of burst oscillations in the cooling tails of thermonuclear bursts from
4U 1728-34 and 4U 1702-429 has been discussed in terms of angular momentum 
conservation of the thermonuclear shell (see \markcite{SJGL}Strohmayer et al 1997; 
\markcite{SM99}Strohmayer \& Markwardt 1999). Expanded at burst onset by the initial 
thermonuclear energy release, the shell spins down due to its larger rotational 
moment of inertia compared to its pre-burst value. As the accreted layer subsequently 
cools its scale height decreases and it comes back into co-rotation with the neutron 
star over $\approx$ 10 seconds. To date the putative initial spin down at burst onset 
has not been observationally confirmed, perhaps due to the radiative diffusion delay, 
on the order of a second, which can hide the oscillations until after the shell has 
expanded and spun down (see, for example, \markcite{B98}Bildsten 1998). We continue to 
search for such a signature, however. Although the initial spin down at burst onset has 
not been seen, the observation of a spin down episode in the {\it tail} of a burst begs 
the question; can it be understood in a similar context, that is, by invoking a second, 
thermal expansion of the burning layers?  The supporting evidence
is the presence of spin down associated with the spectral evidence for an 
additional energy source, the extended tail, as well as the spectral variation observed 
at the time the spin down commenced (see Figure 5). Based on these 
observations we suggest that the spin down began with a second episode of thermonuclear 
energy release, perhaps in a hydrogen-rich layer underlying that responsible for the 
initial instability, and built up over several preceeding bursts. Such a scenario is not
so unlikely based on previous theoretical work (see \markcite{T93}Taam et al. 1993, 
\markcite{F92}Fushiki et al. 1992). The observed rate of spin down, $d_{\nu}^2 = 
-1.01\times 10^{-3}$ s$^{-1}$, interpreted as an increase in the height of the angular 
momentum conserving shell gives $dr/dt \approx (\Delta\nu / 2\Delta T \nu_0)R \approx 
5.25$ m s$^{-1}$, for a neutron star radius of $R = 10$ km. Calculations predict 
increases in the scale height of the bursting layer on the order of 20-30 m during 
thermonuclear flashes (see \markcite{J77}Joss 1977; and \markcite{B95}Bildsten 1995). 
Based on this and the energy evident in the extended tail, the additional expansion of 
about 12 m does not appear overly excessive. If correct this scenario would require 
that the oscillation frequency eventually increase again later in the tail. 
Unfortunately the oscillation dies away before another increase is seen. 

It is interesting to note that bursts from 4U 1636-53 do not appear to show the same 
systematic evolution of the oscillation frequency as is evident in bursts from
4U 1728-34 and 4U 1702-429 (see for example, \markcite{M99b}Miller 1999b; and 
\markcite{SM99}Strohmayer \& Markwardt 1999). In particular, there is no strong 
evidence for an exponential-like recovery that is often 
seen in 4U 1728-34 and 4U 1702-429. Rather, in 4U 1636-53, when the burst oscillation 
frequency reappears after photospheric touchdown in many bursts it appears almost 
immediately at the higher frequency. In the context of a spinning shell this might 
suggest that the shell recouples to the underlying star more quickly than in 4U 1728-34 
or 4U 1702-429. Interestingly, 4U 1636-53 is also the only source to show significant
pulsations at the sub-harmonic of the strongest oscillation frequency, and this has
been interpreted as revealing the presence of a pair of antipodal hot spots (see 
\markcite{M99a}Miller 1999a). These properties may be related and could, for example, 
indicate the presence of a stronger magnetic field in 4U 1636-53 than the other sources.

Another physical process which could alter the observed frequency is related to 
general relativistic (GR) time stretching. If the burst evolution modulates the 
location in radius of the photosphere,
then the rotation at that radius, as seen by a distant observer, is affected by a
redshift such that, $\Delta r /R = (R/r_g)(1-r_g/R)(1-1/(\nu_{h}/\nu_{l})^2)$, where
$\Delta r$ is the change in height of the photosphere, $r_g = 2GM/c^2$ is the 
schwarzchild radius, and $\nu_h/\nu_l$ is the ratio of the highest and the lowest 
observed frequencies.  If this were the sole cause of the frequency changes, then 
it would imply a height change for the photosphere of 
$\approx 120$ m, which is much larger than the increases predicted theoretically
for bursting shells.  Note that this effect works counter to angular momentum 
conservation of the shell, since increasing the height makes the frequency 
higher compared to deeper layers. Since the thicknesses of pre- and post-burst shells 
are on the order of 20 - 50 m, we estimate from the above that the GR correction 
amounts to about 10 - 20\% of the observed frequency change, and, if the angular 
momentum conservation effect is at work, requires a modest {\it increase} in the 
height of the shell over that estimated non-relativistically.

We have reported in detail on the first observations of a spin down in the frequency
of X-ray brightness oscillations in an X-ray burst. We have shown that this event
is coincident with the ocurrence of an extended tail as well as a spectral signature
which both suggest a secondary release of thermonuclear energy in the accreted layer.
It is always difficult to draw conclusions based on a single event, however, 
if the association of spin down episodes with an extended X-ray tail can be 
confirmed in additional bursts this will provide strong evidence in support of the
hypothesis that angular momentum conservation of the thermonuclear shell is 
responsible for the observed frequency variations during bursts. The combination
of spectral and timing studies during bursts with oscillations can then give us a
unique new probe of the physics of thermonuclear burning on neutron stars. 

\acknowledgements

We thank Craig Markwardt and Jean Swank for many helpful discussions. 

\vfill\eject

\vfill\eject

\section{Figure Captions}

\noindent Figure 1: Dynamic variability spectra for bursts A (top) and B (bottom)
from 4U 1636-53. Shown are contours of constant power $Z_1^2$ computed from 2 s 
intervals with a new interval every 0.25 s. The countrate vs. time in the PCA is 
also shown. For burst A, the two segment frequency evolution model is shown as the 
solid curve. The best fitting model parameters were; $\nu_0 = 580.70$ Hz, 
$d_{\nu}^1 = 9.0\times 10^{-5}$ s$^{-1}$, $d_{\nu}^2 = -1.0 \times 10^{-3}$ s$^{-1}$, 
and $t_b = 10.86$ s. 

\vskip 10pt

\noindent Figure 2:  A plot of $Z_1^2$ vs. frequency parameter $\nu_0$ for the spin 
down burst (burst A).The solid curve shows the result for the best fitting two-segment 
frequency evolution model. The dashed curve was produced assuming no frequency 
evolution. 

\vskip 10pt

\noindent Figure 3: 2 - 20 keV light curves for bursts A (solid) and B (dashed) from 
4U 1636-53. Notice the long, extended tail in burst A. The pre-burst and peak countrates
were virtually the same for both bursts. The bursts were aligned in time to facilitate 
direct comparison. 

\vskip 10pt

\noindent Figure 4: Results of spectral evolution analysis for bursts A and B.
The top panel shows evolution of the bolometric flux deduced from the best 
fitting black body parameters for bursts A (solid) and B (dashed). Note the long
tail on burst A, and that the peak fluxes are consistent. The bottom panel shows
the evolution of the black body temperature $kT$ for bursts A (solid) and B (dashed).
The initial drop in $kT$ followed by an increase is a signature of radiation
driven photospheric expansion. The dotted vertical line marks $t_b$, the break time
which marks the onset of spin down (see discussion in text).

\vskip 10 pt

\noindent Figure 5: Evolution of the black body temperature $kT$ (dashed) and inferred
radius $R_{BB}$ (solid) for the spin down burst (burst A). The dotted vertical line 
marks $t_b$, the break time which marks the onset of spin down (see discussion in text).
The burst begins with an episode of photospheric radius expansion, marked by the 
simultaneous decrease in temperature and increase in radius. Notice the secondary 
signature of a weak radius expansion event near time $t_b$ (dotted vertical line). 

\vfill\eject

\begin{figure*}[htb] 
\centerline{\epsfig{file=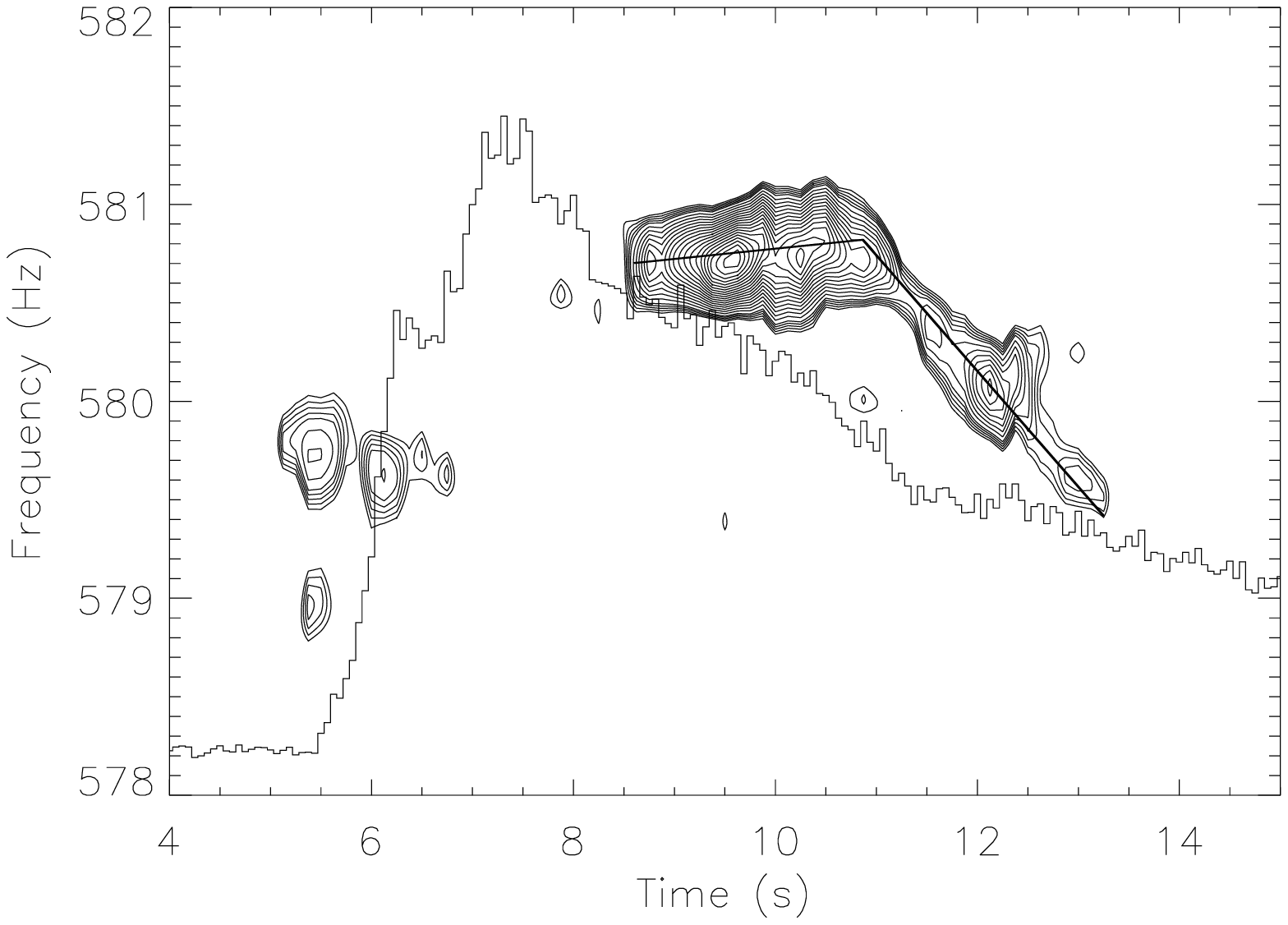,height=6.0in,width=6.0in}}
\vspace{10pt}
\caption{Figure 1 (top)}
\end{figure*}

\vfill\eject

\begin{figure*}[htb] 
\centerline{\epsfig{file=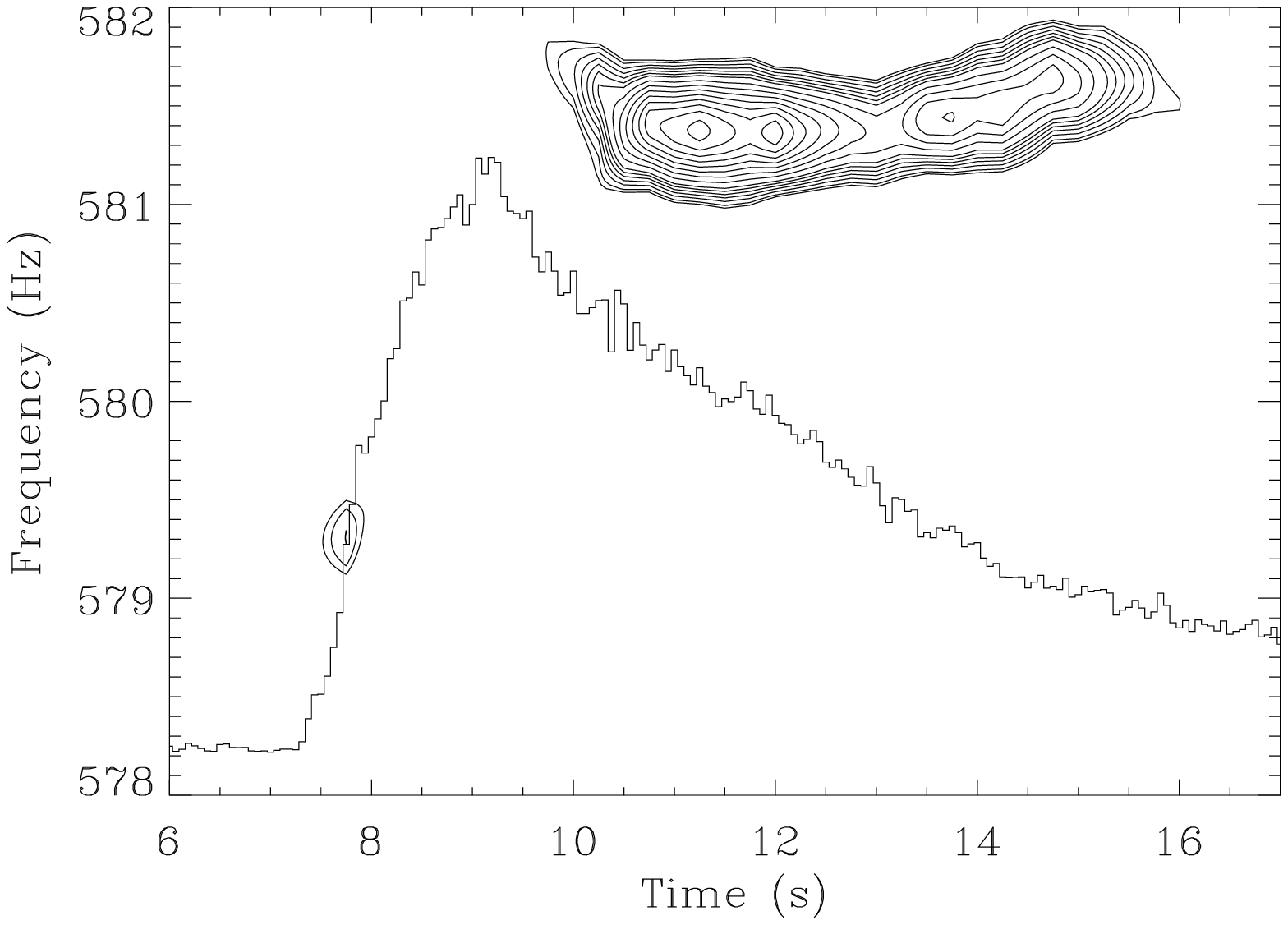,height=6.0in,width=6.0in}}
\vspace{10pt}
\caption{Figure 1 (bottom)}
\end{figure*}

\vfill\eject

\begin{figure*}[htb] 
\centerline{\epsfig{file=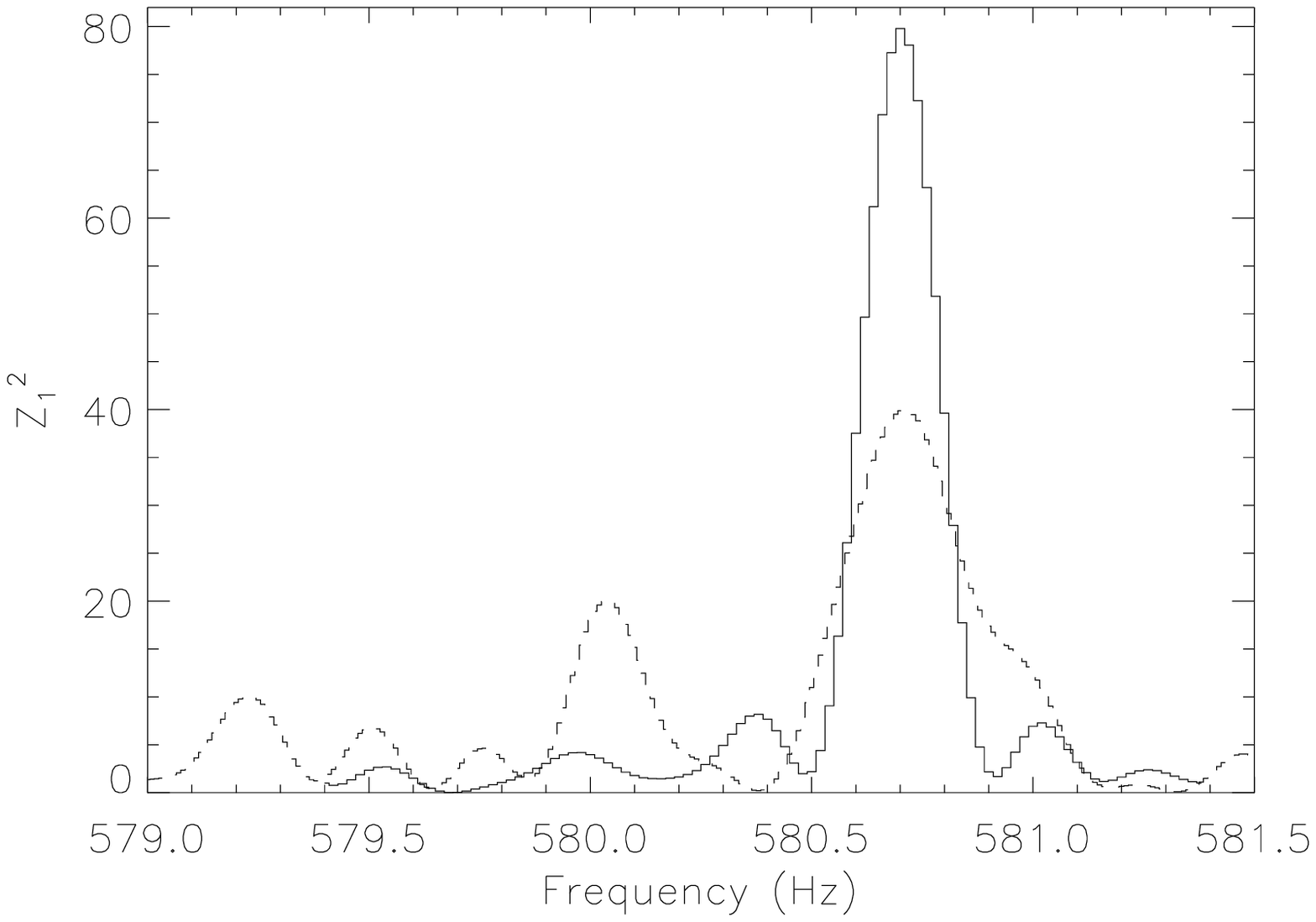,height=6.0in,width=6.0in}}
\vspace{10pt}
\caption{Figure 2}
\end{figure*}

\vfill\eject

\begin{figure*}[htb] 
\centerline{\epsfig{file=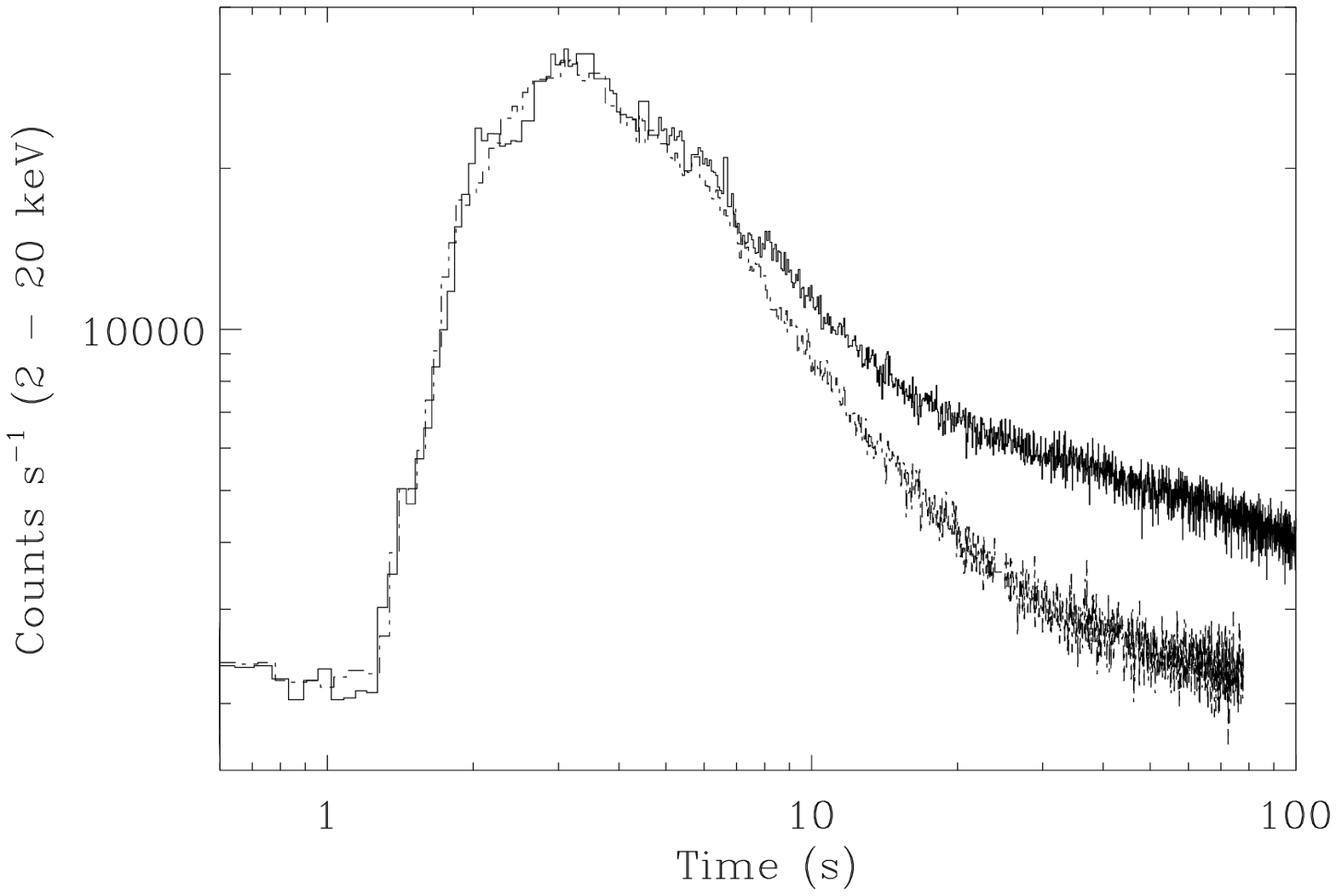,height=6.0in,width=6.0in}}
\vspace{10pt}
\caption{Figure 3}
\end{figure*}

\vfill\eject

\begin{figure*}[htb] 
\centerline{\epsfig{file=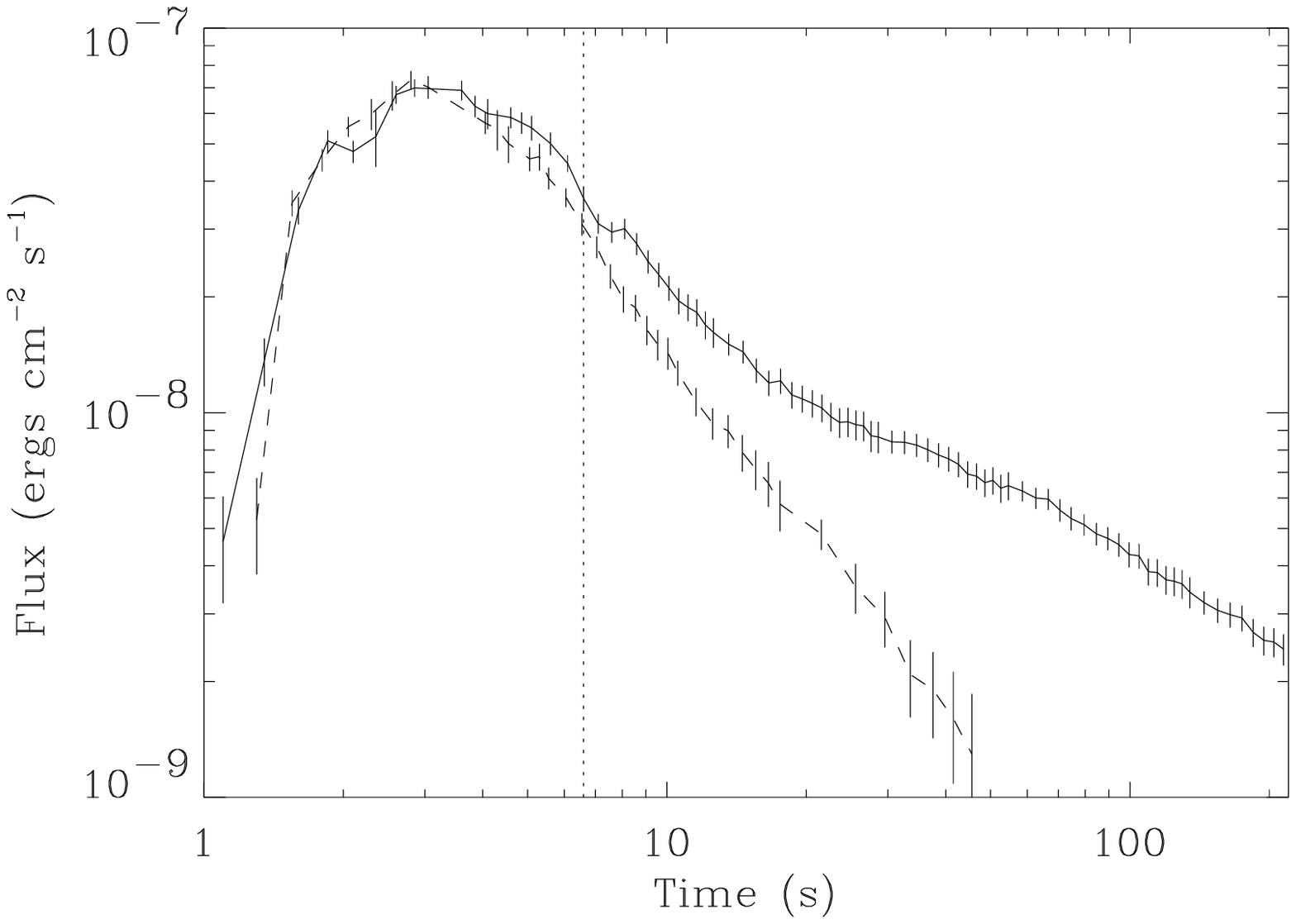,height=6.0in,width=6.0in}}
\vspace{10pt}
\caption{Figure 4a}
\end{figure*}

\vfill\eject

\begin{figure*}[htb] 
\centerline{\epsfig{file=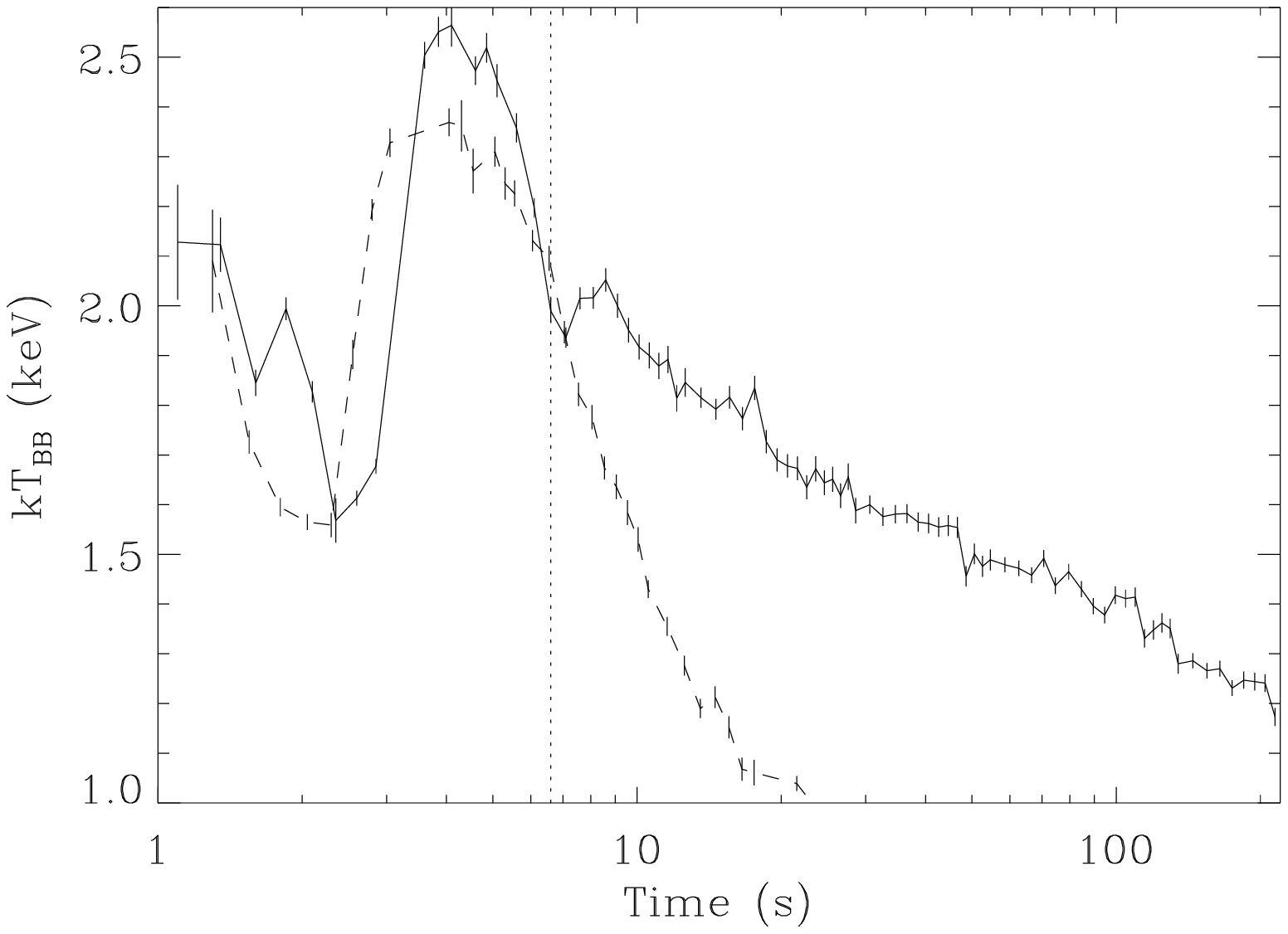,height=6.0in,width=6.0in}}
\vspace{10pt}
\caption{Figure 4b}
\end{figure*}

\vfill\eject

\begin{figure*}[htb] 
\centerline{\epsfig{file=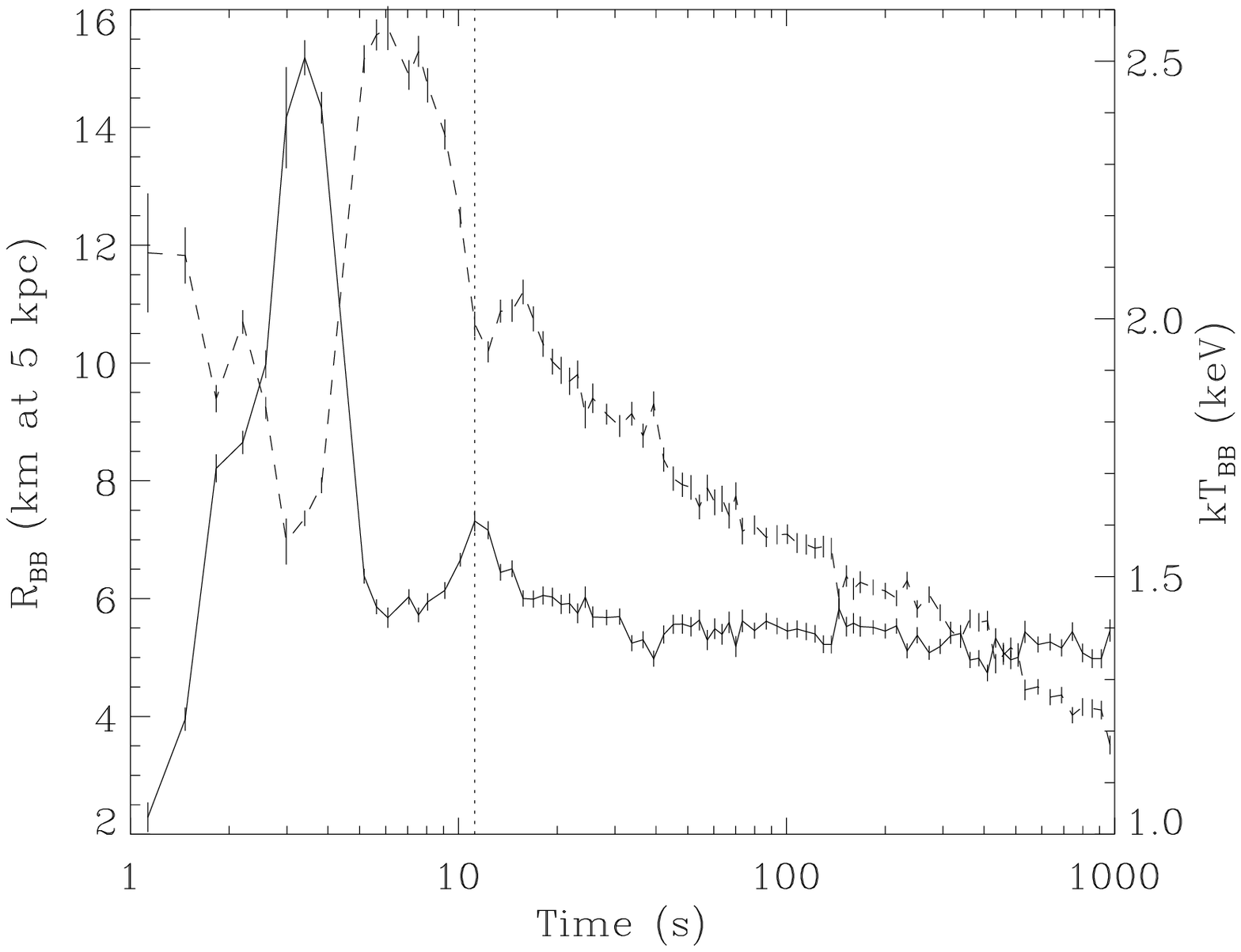,height=5.0in,width=5.0in}}
\vspace{10pt}
\caption{Figure 5}
\end{figure*}

\end{document}